# Authentication System Securing Index of Image using SVD and ECC

[1] Tahina Ezéchiel Rakotondraina, [2] Dr. Henri Bruno Razafindradina

1 Department of Telecommunication, High School Polytechnic of Antananarivo, University of Antananarivo
Antananarivo, Ankatso BP 1500, Madagascar

2 Department of Telecommunication, High Institute of Technology
Diego Suarez, Madagascar

**Abstract**
This paper presents a new approach to securing information stored in a database. It has three components including: an operation for indexing images using Singular Value Decomposition (SVD), which will constitute the reference images, asymmetric encryption operation using Elliptic Curve Cryptosystem (ECC), aiming to make confidential these reference images stored and a technique for comparing these images to a query image using the Euclidian Distance..

## 1. Introduction

In this approach, our goal is to test a new authentication system, at the same time we secure the information in a database by only storing the information representing. This way gives us more free memory at the Data base table.
The method is divided into two parts, namely:

First, we will make treatments on information stored in the database, the information or specifically the fingerprinting image is undergoing an indexing operation in the transformed domain using Singular Value Decomposition (SVD). This operation aims to decompose the image into three matrixes which represent each other a specific detail.

Second, after decomposing the image into three components according to the method chosen, we apply asymmetric encryption using Elliptic Curve Cryptosystem (ECC), on the index of the image with the information representative. The objective of encryption is to provide confidentiality and integrity of the index. This part encrypted will be stored in the database until the comparison with the query image.Third, to authenticate, the user uses electronic or optical device to obtain a fingerprint. After collection, the query image undergoes an identical indexing operations treatment like the stored images. To enable authentication, we decrypt the data in the database and compared with the index of the query image. The comparison is performed by calculating the similarity of the two indexes by using the Euclidean distance.

## 2. Proposed approach

The flow diagram is given by the following figure:

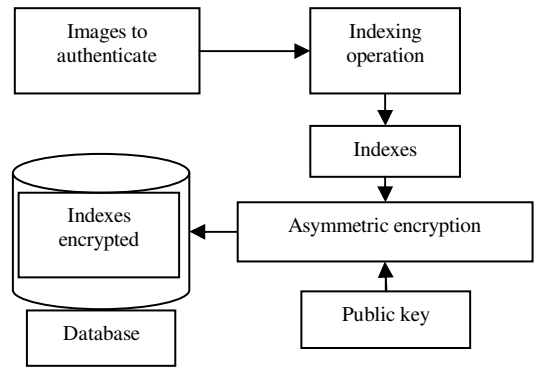

Fig. 1 Securing Scheme

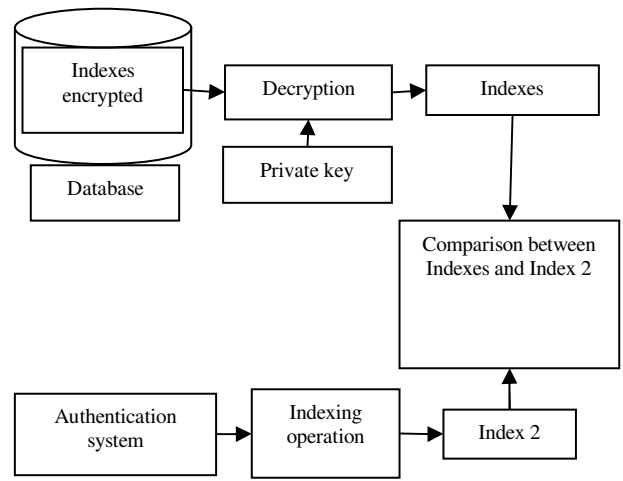

Fig. 2 Comparison Scheme

## 3. Results and interpretations

With SVD indexing method, we obtain three matrixes: $A = U * S * V^T$, such that the matrix S represent more the image and what is it that we encrypt.

Fingerprints used in the experiment are:

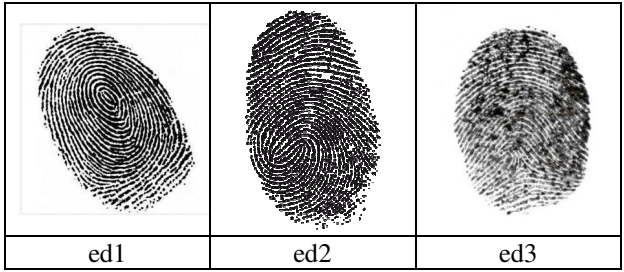

ed1   ed2   ed3

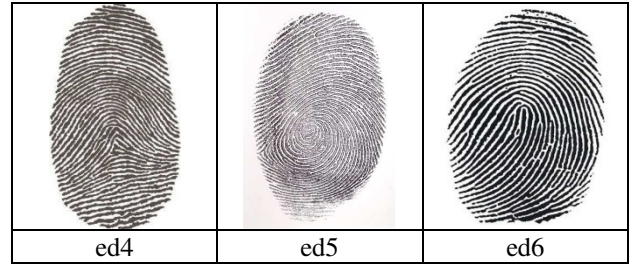

ed4   ed5   ed6

Fig. 3 Fingerprints use

We have the following table (Table 1) which shows the similarity between the query images and index images stored in the database (after decryption). Series of test are carried out on an Intel Pentium Dual Core 2.2 GHz with 3072 MB of RAM.

Table 1: Index similarity values

|     | *ed1*     | *ed2*     | *ed3*     | *ed4*     | *ed5*     | *ed6*     |
| --- | --------- | --------- | --------- | --------- | --------- | --------- |
| *ed1* | 0 | 6.7311 e+3 | 1.4921 e+3 | 1.0350 e+3 | 952.6909 | 358.9680 |
| *ed2* | 6.7311 e+3 | 0 | 7.5955 e+3 | 6.9186 e+3 | 6.9467 e+3 | 6.8422 e+3 |
| *ed3* | 1.4921 e+3 | 7.5955 e+3 | 0 | 2.1550 e+3 | 2.1039 e+3 | 1.5101 e+3 |
| *ed4* | 1.0350 e+3 | 6.9186 e+3 | 2.1550 e+3 | 0 | 650.5721 | 933.5138 |
| *ed5* | 952.6909 | 6.9467 e+3 | 2.1039 e+3 | 650.5721 | 0 | 973.9959 |
| *ed6* | 358.9680 | 6.8422 e+3 | 1.5101 e+3 | 933.5138 | 973.9959 | 0 |

Table 1 above shows the effectiveness of our approach because the probability of encountering two identical fingerprints of two different people tends to zero. It should be noted that our approach is invariant against geometric transformations such as rotation, scaling and translation of the query image because we used the values of the histogram of the indexed image. We can also see the execution time (in seconds) of the program: the indexing operation, encryption and decryption in the following table:

Table 02: computation speed of the indexing, encryption, decryption programs

| *Indexing time* | *Encryption time* | *Decryption time* |
| --- | --- | --- |
| 0.0828 [s] | 0.0346 [s] | 0.0488 [s] |

Comparing our indexing method by another method such as Principal Component Analysis (PCA), PCA has a large advantage on the computation speed compared to the indexing method by SVD, in the fact of, SVD uses additional operations when extracting the energy level of the image. But, the SVD method has high efficiency on the accuracy of signatures, that is to say the maximum energy representing information.

## 4. Conclusion

Our approach is based on how to represent the index of a fingerprint by the SVD indexing method. The problem of limited storage capacity is determined from the fact that only the representative part of the image is stored but not the whole picture. We also used a very powerful cryptosystem, which is asymmetric cryptography based on elliptic curves (ECC). It allows a level of security similar to the RSA cryptosystem, which is still the most used, using an encryption key of 1024 bits, against key seven times shorter in ECC. Finally, the evaluation is done by calculating the similarity between the query image and reference images decrypted.

**Acknowledgments**

Authors thank French Cooperation through the MADES project for its Sponsor and financial support.

**Tahina E. Rakotondraina** was born in Antsirabe, Madagascar on 1984. He received his M.S. degrees in Information Theory and Cryptography in 2010 at University of Antananarivo (Madagascar). He works as a Teacher assistant and a Ph.D. student at High School Polytechnic of Antananarivo. His currents research interests include Cryptography, multimedia, Information Hiding, VOIP. He is a co-author of two papers published in international journal.

**Henri B. Razafindradina** was born in Fianarantsoa, Madagascar, on 1978. He received, respectively, his M.S degree and Ph.D. in Computer Science and Information Engineering in 2005 and 2008. He served since 2010 as a professor at High Institute of Technology Diego Suarez, became an assistant lecturer in 2011. His current research interests include Images compression, multimedia, computer vision, informationHiding.